\documentclass[conference]{IEEEtran}
\IEEEoverridecommandlockouts
\usepackage{cite}
\usepackage{amsmath,amssymb,amsfonts}
\usepackage{algorithmic}
\usepackage{graphicx}
\usepackage{textcomp}
\usepackage{multirow}
\usepackage[font=sf]{caption}
\usepackage[font=sf]{subcaption}
\usepackage[table,xcdraw]{xcolor}
\def\BibTeX{{\rm B\kern-.05em{\sc i\kern-.025em b}\kern-.08em
    T\kern-.1667em\lower.7ex\hbox{E}\kern-.125emX}}

\usepackage{booktabs} 
\usepackage{listings}
\usepackage{tikz}
\usetikzlibrary{positioning}

\usetikzlibrary{quantikz2}
\usepackage{multicol}
\usepackage[scale=0.90]{plex-sans}
\usepackage[scale=0.90]{plex-mono}
\usepackage{braket}
\usepackage{cuted}

\usepackage{pgfplots}

\definecolor{codegreen}{rgb}{0,0.6,0}
\definecolor{codegray}{rgb}{0.5,0.5,0.5}
\definecolor{codepurple}{rgb}{0.58,0,0.82}
\definecolor{backcolour}{rgb}{0.95,0.95,0.92}

\lstdefinestyle{mystyle}{
    backgroundcolor=\color{backcolour},   
    commentstyle=\color{codegreen},
    keywordstyle=\color{magenta},
    numberstyle=\tiny\color{codegray},
    stringstyle=\color{codepurple},
    basicstyle=\ttfamily\footnotesize,
    breakatwhitespace=false,         
    breaklines=true,                 
    captionpos=b,                    
    keepspaces=true,                 
    numbers=left,                    
    numbersep=5pt,                  
    showspaces=false,                
    showstringspaces=false,
    showtabs=false,                  
    tabsize=2
}

\makeatletter
\renewenvironment{table}%
  {\renewcommand\familydefault\sfdefault
   \@float{table}}
  {\end@float}
\makeatother

\begin{document}

\setlength{\textfloatsep}{3pt}
\title{
    \footnotesize Submitted as New Ideas and Emergent Results (NIER) paper at QCE24.
    \\\huge \textbf{XGSwap}: eXtreme
Gradient boosting Swap \\for Routing in NISQ Devices

}

\author{
 \IEEEauthorblockN{Jean-Baptiste Waring}
 \IEEEauthorblockA{\textit{\footnotesize Dept. of Electrical \& Computer Engineering} \\
 \textit{Concordia University}\\
 Montréal, Québec \\
 j\_warin@live.concordia.ca}
 \and
 \IEEEauthorblockN{Christophe Pere}
 \IEEEauthorblockA{\textit{\footnotesize Dept. of Computer Science \& Software Engineering} \\
 \textit{INTRIQ, Laval University,}\\
 Québec, QC, Canada \\
 PINQ2, \\
 christophe.pere.1@ulaval.ca}
 \and
 \IEEEauthorblockN{Sébastien Le Beux}
 \IEEEauthorblockA{\textit{\footnotesize Dept. of Electrical \& Computer Engineering} \\
 \textit{Concordia University}\\
 Montréal, Québec \\
 sebastien.lebeux@concordia.ca}
\vspace{-1cm}
}

\maketitle

        
            \begin{abstract}
\mdseries
In the current landscape of noisy intermediate-scale quantum (NISQ) computing, the inherent noise presents significant challenges to achieving high-fidelity long-range entanglement.
Furthermore, this challenge is amplified by the limited connectivity of current superconducting devices, necessitating state permutations to establish long-distance entanglement.
Traditionally, graph methods are used to satisfy the coupling constraints of a given architecture by routing states along the shortest undirected path between qubits.
\textbf{In this work, we introduce a gradient boosting machine learning model to predict the fidelity of alternative--potentially longer--routing paths to improve fidelity. This model was trained on 4050 random CNOT gates ranging in length from 2 to 100+ qubits. The experiments were all executed on \texttt{ibm\_quebec}, a 127-qubit IBM Quantum System One. Through more than 200+ tests run on actual hardware, our model successfully identified higher fidelity paths in approximately 24\% of cases.} 
\end{abstract}


\section{Introduction}

The advancement of quantum computing heralds a new era of computational capabilities, poised to address challenges beyond the reach of traditional computing methods--such as factoring very large prime numbers in polynomial time \cite{shor_polynomial-time_1999}. With each passing year, the field of quantum computing has achieved notable progress with claims of utility \cite{kim_evidence_2023}. Yet, with the increasing sophistication of quantum hardware and the complexity of quantum algorithms, there emerges a critical demand for more effective methods for routing and circuit layout optimization.
Quantum algorithms' efficacy can be deeply influenced by the choice of routing and layout strategies. For quantum architectures restricted to local qubit interactions, state permutations are required to comply with coupling restrictions, given that duplicating a quantum state is prohibited by the no-cloning theorem \cite{wootters_single_1982}. This requirement often results in an increase in circuit depth proportional to the number of state permutations, which in turn can significantly limit the fidelity of large circuits operated on Noisy Intermediate Scale Quantum (NISQ) devices \cite{preskill_quantum_2018}, characterized by their limited qubit connectivity and high noise levels. Therefore, enhancing the performance of quantum computations requires a sophisticated strategy that not only considers the inherent noise in the system but also tailors to the unique noise profiles and connectivity limitations of each quantum device.

\textbf{In this paper, we introduce XGSwap (eXtreme Gradient boosting Swap), a model based on gradient boosting to estimate the fidelity of long-range CNOT gates used in quantum circuit routing.}

Our model, developed using a dataset of over 4000 random CNOT gates of varied lengths, is utilized to address the coupling constraints of a given architecture by identifying and selecting the path with the highest predicted fidelity.

This paper is organized in the following manner: Section II provides background information on routing, process tomography, and gradient boosting techniques. Section III describes our methods for data collection and model integration. Section IV addresses the training and validation processes. Finally, Section V showcases the results of our method when applied to random entangling gates and compares these with Qiskit's \cite{treinish_qiskitqiskit-metapackage_2023} default behavior. 


\section{Background \& Related Work}

In this section, we provide an overview of the key concepts and techniques that underpin our work. We begin by introducing the layout \& routing problem, a critical challenge in addressing the coupling constraints of NISQ devices. We then discuss quantum process tomography and fidelity, which we use as our quality metric. Then, we provide an overview of gradient boosting, the machine learning technique we use to predict the fidelity of CNOT gates. Finally, we discuss related work in quantum circuit mapping and machine learning approaches to the problem.

%
%
\subsection{Quantum Computing and Routing}
\begin{figure}[h!]
    \centering
    \begin{tabular}{c}\begin{quantikz}
                \qw & \ctrl{1} & \qw  \\
                \qw & \targ{} & \qw 
            \end{quantikz} = $\begin{bmatrix}
                1 & 0 & 0 & 0 \\
                0 & 1 & 0 & 0 \\
                0 & 0 & 0 & 1 \\
                0 & 0 & 1 & 0
            \end{bmatrix}$
    \end{tabular}
    \caption{Quantum Circuit and matrix representation of a CNOT gate. \cite{nielsen_quantum_2010}}
    \label{fig:cnot-gate}
\end{figure}
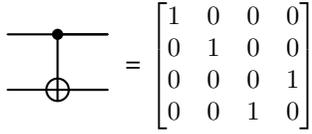

Quantum computing exploits the principles of quantum mechanics to process information in ways that are profoundly different from classical computing. A key operation in many quantum algorithms is the controlled-NOT (CNOT) gate (Fig. \ref{fig:cnot-gate}), which is essential for entangling qubits \cite{nielsen_quantum_2010}.

\begin{figure}[h!]
    \centering
    \includegraphics[width=\linewidth]{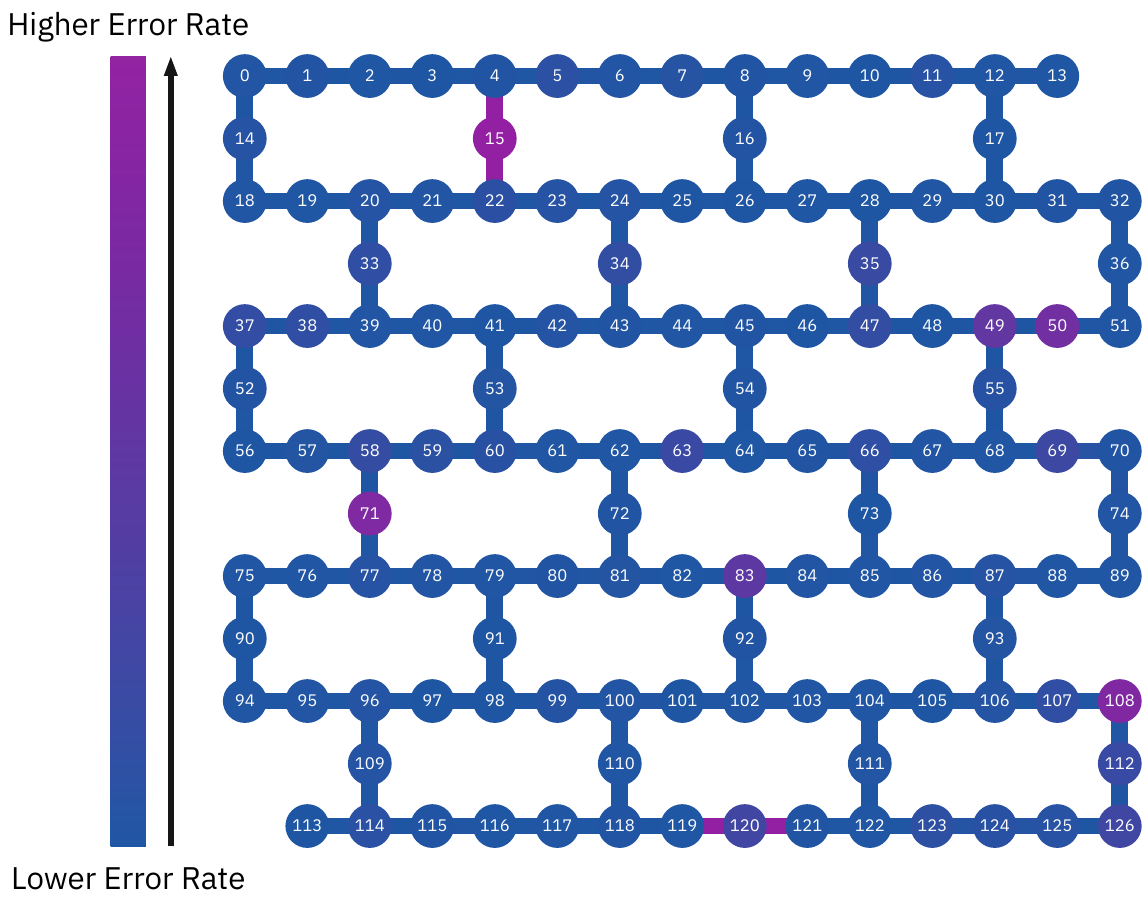}
    \caption{Coupling map of \texttt{ibm\_quebec}, a 127-qubit Eagle Quantum Processor representing the connectivity between qubits. The color of node and edges represent readout and entangling error rates, respectively.}
    \label{fig:coupling-map-quebec}
\end{figure}

Quantum processors typically do not have all-to-all connectivity, meaning not all qubits can directly interact with each other \cite{de_micheli_advances_2022}. This architectural limitation necessitates effective routing strategies to rearrange quantum states such that required interactions occur in compliance with the hardware's physical connectivity constraints. We represent the possible interactions between qubits using a coupling map (Fig. \ref{fig:coupling-map-quebec}) which usually includes error-rates using a color encoding.

Routing in quantum computing involves adapting a quantum circuit to fit the specific topology --or coupling constraints--of a quantum processor. The goal is to minimize the overhead introduced by additional gates needed to move quantum information across the processor, which can significantly impact the overall fidelity and performance of quantum operations. The SABRE (SWAP-based BidiREctional heuristic) \cite{li_tackling_2019} is a state-of-the-art method in quantum circuit routing (see \cite{noauthor_sabreswap_nodate} for Qiskit impl.). This approach prioritizes reducing the number of additional SWAP gates needed to adhere to the coupling constraints, thereby optimizing the layout for improved computational efficiency and reduced errors. However, SABRE is not itself a routing primitive and, as such, tries to minimize the total number of SWAP gates but does not directly consider their fidelity. 

In summary, effective routing is paramount in quantum computing as it directly influences the feasibility and efficiency of executing quantum circuits on real hardware.

%
%
\subsection{Quantum Process Tomography}

Quantum Process Tomography (QPT) is a technique used to characterize the dynamics of a quantum system \cite{nielsen_quantum_2010}.
It involves performing a set of measurements on a quantum system to reconstruct the complete quantum state and understand how a quantum process transforms that state. Specifically, we utilize process tomography to determine the process fidelity, which quantifies how closely the experimental process matches the ideal quantum operation. This process fidelity is then converted to gate fidelity, providing a more direct measure of the effectiveness of quantum gates such as the CNOT. Gate fidelity is essential for assessing the performance of quantum gates in practical settings, where achieving high fidelity is indicative of operational accuracy and reliability in quantum computing.

\subsection{Gradient Boosting}

XGBoost or eXtreme Gradient Boosting is a tree-based algorithm\cite{chen_xgboost_2016}. At its core, XGBoost belongs to the family of ensemble learning techniques, specifically boosting algorithms. Boosting, a sequential ensemble method, aims to combine the predictive capabilities of multiple weak learners--typically decision trees in the case of XGBoost--to create a robust and accurate predictive model. Unlike traditional bagging methods--which build models independently and then average their predictions--, boosting algorithms construct a series of models iteratively, with each subsequent model focusing on the errors made by its predecessors, thereby continuously improving predictive performance.

XGBoost employs decision trees as its base learners. Decision trees partition the feature space into smaller regions and assign a predictive value to each region. In the context of XGBoost, decision trees are often shallow, with a limited number of levels, also known as weak learners.


\subsection{Related Work}
Quantum circuit mapping, an NP-hard problem \cite{siraichi_qubit_2018, botea_complexity_2018}, has been studied using classical approaches and heuristics in the literature such as in \cite{wille_exact_2014, wille_mapping_2019, li_tackling_2019}. Additionally, quantum teleportation has been explored as a potential routing primitive in \cite{hillmich_exploiting_2021} and its application for teleporting gates further investigated in \cite{baumer_efficient_2023}, focusing on refining the routing primitive rather than optimizing the path it takes.
Furthermore, machine learning algorithms have also been proposed to address the circuit mapping problem. In \cite{pozzi_using_2022}, a modified deep Q-learning\footnote{A convolutional neural network framework for learning a function $Q(s, a)$ that represents the quality of being in a state $s$ and taking action $a$. Here the $Q$ is for \textit{quality}, not \textit{quantum}.} \cite{mnih_playing_2013} formulation that selects routing actions from a combinatorial space is used. 
Deep neural network approaches have also been proposed such as by \cite{acampora_deep_2021}. However, these approaches require extensive training since they rely on a model to produce the mapping itself. In contrast, our approach, in essence, provides a cost function based on path fidelity predictions to existing mapping heuristics and does not require extensive training. Moreover, \cite{vadali_quantum_2023} has explored the prediction of quantum circuit fidelity, yet solely employing simulated noise models that do not accurately reflect actual hardware conditions, and utilizing Convolutional Neural Networks (CNNs) for this purpose.

%
%
\section{Method}
The principal aim of our approach is to demonstrate the effectiveness of gradient boosting techniques, specifically XGBoost \cite{chen_xgboost_2016}, in predicting the gate fidelity of CNOT gates within current quantum systems. 
Our end-goal is to improve routing strategies by selecting paths with the highest predicted fidelity rather than simply the shortest undirected paths in the coupling map. This optimized routing strategy is intended to improve the performance and reliability of computations on quantum systems, especially those with heterogeneous error characteristics and time-dependent calibration states.

%
%
\subsection{Method Overview}
\begin{figure}[h!]
    \centering
    \includegraphics[width=\linewidth]{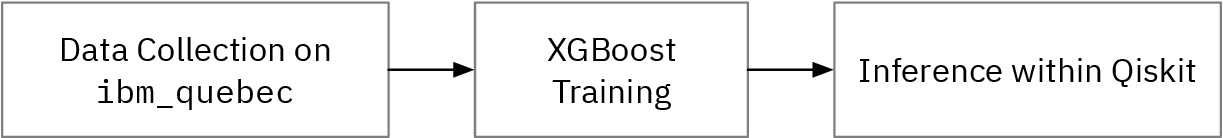}
    \caption{High-level method overview.}
    \label{fig:method-overview}
\end{figure}
Our method consists of three main stages (Fig. \ref{fig:method-overview}): data collection, model training, and inference within Qiskit.
The data collection stage consists of running benchmarks of random CNOT gates on a real quantum backend. The data collected includes the fidelity of the CNOT gates, the paths taken between the source and target qubits, and the timestamp of each execution. This data is then used to train a gradient-boosting model to predict the fidelity of CNOT gates based on the path taken between the source and target qubits. The trained model is then integrated into Qiskit using a custom transpiler pass. This pass is either executed as the routing primitive to the SabreLayout \cite{noauthor_sabrelayout_nodate} (layout stage) or as a standalone routing path in the routing stage. Since our validation involves specifying the initial layout to set the control and target qubits, no additional layout pass is required and thus, we do not consider the interplay with SABRE \cite{li_tackling_2019} in this work.

\subsection{Data Collection}
\label{sec:data-collection}

\begin{figure}[h!]
    \centering
    \includegraphics[width=0.8\linewidth]{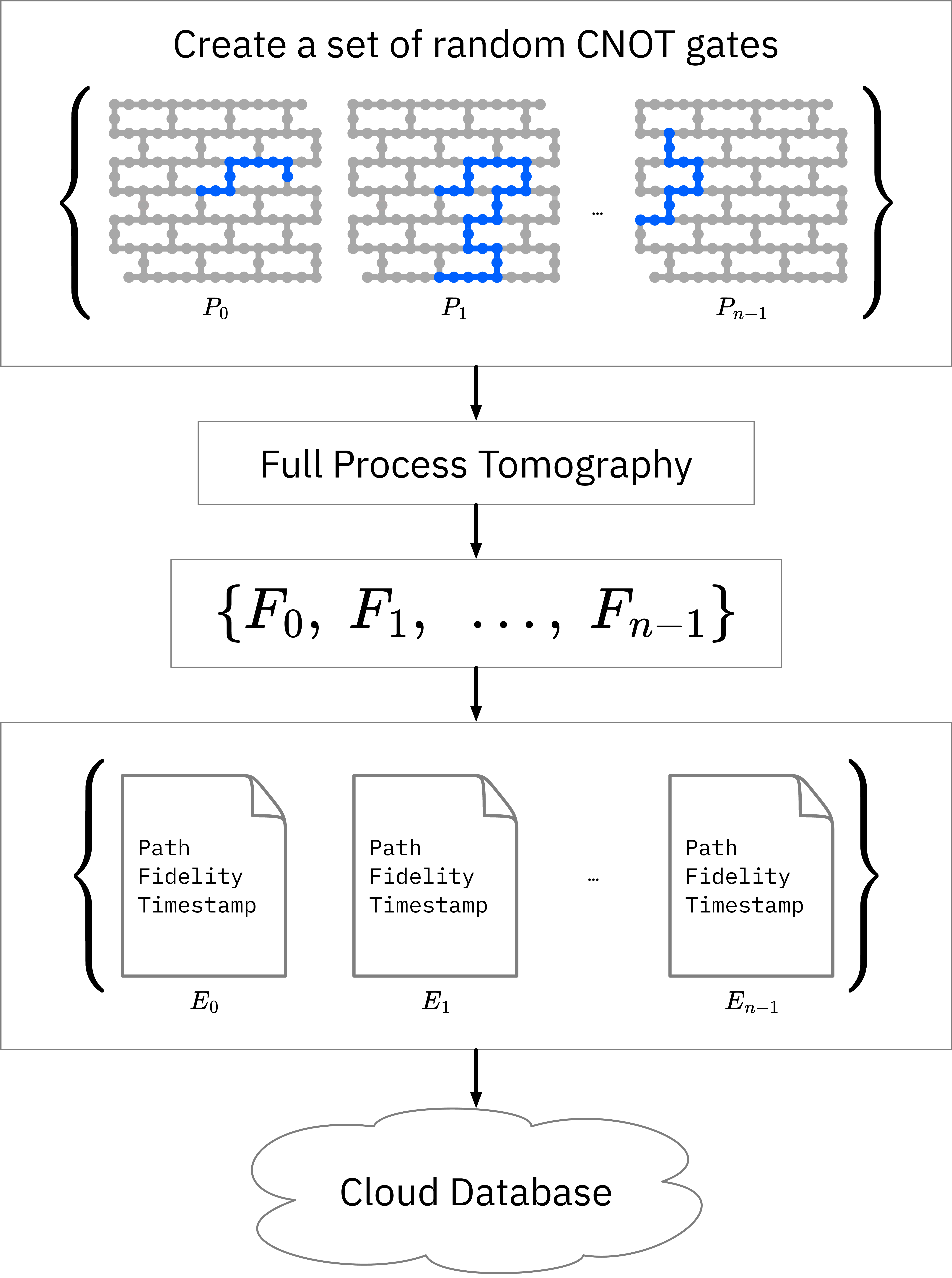}
    \caption{Data collection method, performed between October 2023 and March 2024 on the 127-qubit \texttt{ibm\_quebec} computer.}
    \label{fig:method-data-collection}
\end{figure}

\begin{figure}[h!]
        \centering
        \includegraphics[width=0.8\linewidth]{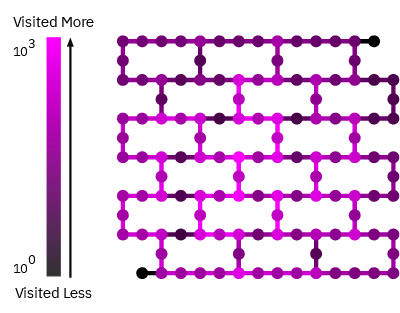}
        \caption{Visualization of qubits and connections visitations frequencies over 4050 experiments for the collected dataset on \texttt{ibm\_quebec}. Node and edge colors indicate visitation counts ranging from a minimum of 43 to a maximum of 2948 for qubits, and from 43 to 2232 for edges, with lighter shades representing higher frequencies.}
        \label{fig:visitation-probability}
    \end{figure}

To develop the dataset needed for training our machine learning model to predict fidelity (Fig. \ref{fig:method-data-collection}), we execute CNOT gates across a range of lengths between randomly chosen source and target qubits. Since CNOT gates involve two qubits, conducting full process tomography requires running $4^N3^N = 144$ circuits to accurately determine the process fidelity. This is further converted to gate fidelity using:
\begin{equation}
    \label{eq:gate-fidelity}
    F_{\text{gate}} = \frac{dF_{\text{process}} + 1}{d+1} = \frac{3 F_{\text{process}} + 1}{4}
\end{equation}
This data, including the specific paths taken between the control and target qubits (detailing the \textit{visited} qubits) and the timestamp of each execution, is then stored in a purpose-built database. Storing this data allows us to correlate the observed fidelity and the path used with the device's calibration data at the time of each test, which would be the model's intended behaviour after training. Some key statistics of the collected dataset are summarized in Tab. \ref{tab:data-collection-summary}. A visualization of the coverage of our random experiments is shown Fig. \ref{fig:visitation-probability}. The full dataset will be made publicly available upon publication.

\begin{table}[h!]
    \caption{Summary of data collection. The dataset was collected between October 2023 and March 2024 on the 127-qubit \texttt{ibm\_quebec} computer.}
    \centering
    \scalebox{0.85}{\begin{tabular}{@{}cccc@{}}
        \toprule
        Number of Experiments & Min Length & Max Length & Average Length \\ \midrule
        4050                  & 2          & 109        & 53.46          \\ \bottomrule
        \end{tabular}}
    \label{tab:data-collection-summary}
    \end{table}

\subsection{Qiskit Integration}

\begin{figure}[h!]
    \centering
    \includegraphics[width=0.8\linewidth]{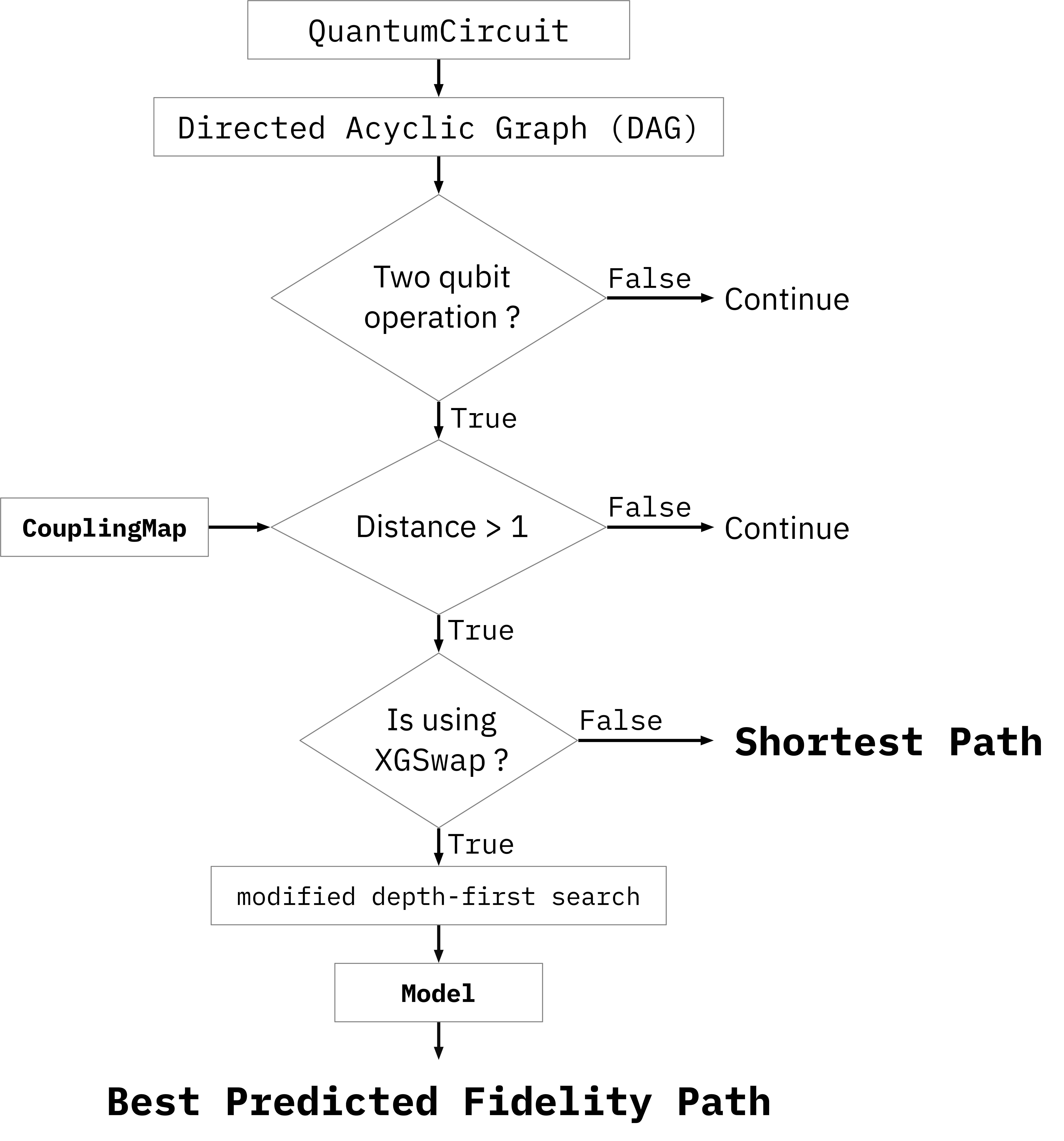}
    \caption{Using our model within Qiskit.}
    \label{fig:inference-diagram}
\end{figure}

One practical application of our method is its integration into the Qiskit framework as a transpiler plugin\footnote{Dataset and Plugin code will be made publicly available upon publication.} (Fig. \ref{fig:inference-diagram}). This custom plugin follows a modular architecture, seamlessly integrating into the transpilation process, particularly during the layout and routing stages. It can be enabled by users to augment the native Qiskit transpiler's capabilities.

The selection of this plugin by the user is straightforward: it can be specified when using the \texttt{transpile} function:
\begin{lstlisting}[language=Python]
transpile(circuit, backend=backend, optimization_level=3, routing_method='xg_swap')
\end{lstlisting}
Upon invocation, our plugin overrides the default shortest-path routing algorithm with our fidelity-optimized predictive strategy. Instead of merely considering the physical distance between qubits on the hardware's coupling map, it employs the fidelity predictions provided by our trained XGBoost model to guide the path selection process.
For a given two-qubit gate, if the coupling distance between the two qubits is more than one, the plugin uses a modified depth-first search algorithm to find a collection of paths between the source and target qubits. It then ranks these paths based on the predicted fidelity of the CNOT gate on each path. The path with the highest predicted fidelity is selected for the gate execution. This process is repeated for each two-qubit gate in the circuit, ensuring that the entire circuit is transpiled with fidelity-optimized routing.

\section{Training \& Validation}
We structure our approach into three main stages: constructing the input vector, sampling data for training and validation, and evaluating the machine learning model's predictions post-training using relevant metrics.

\subsection{Data Preparation}
\label{sec:train-data-preparation}
\begin{figure}
    \centering
    \includegraphics[width=0.8\linewidth]{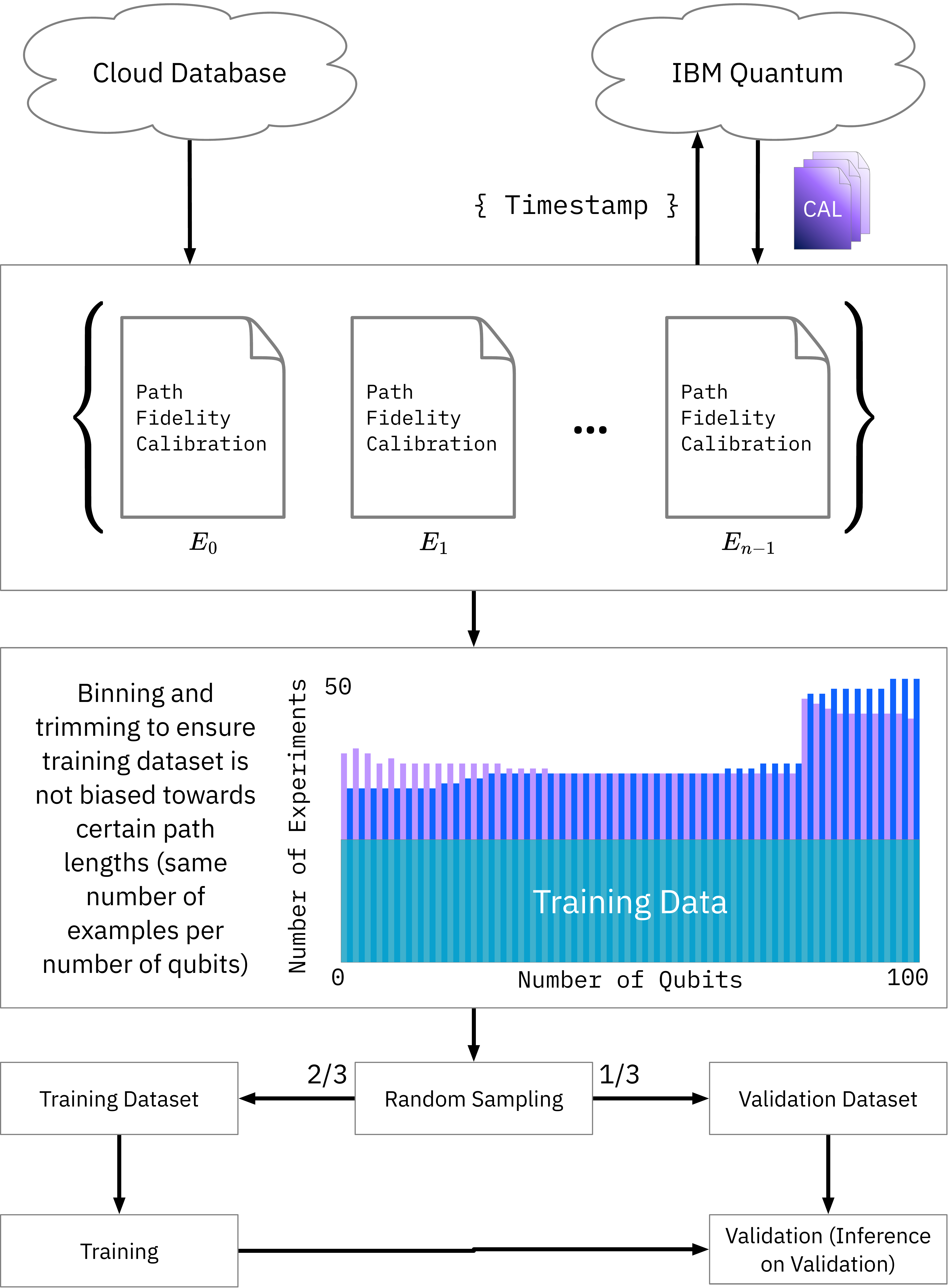}
    \caption{Data Preparation Flow}
    \label{fig:data-preparation-flow}
\end{figure}

We compiled a comprehensive dataset using the method described in Sec. \ref{sec:data-collection}. We first aggregate the data from all experiments to prepare the training and validation datasets (Fig. \ref{fig:data-preparation-flow}). We then exchange timestamps for calibration data. We then bin the data by path length and make a random sample of $2/3$ of the minimum number of experiments--or \textit{examples}-- for each path length, hereby ensuring no bias towards certain path lengths. The remainder of the data can then be used as a validation dataset (the model does not learn during inference, and thus, it does not matter if a skewed dataset is presented).

\begin{figure}[h!]
    \centering
    \includegraphics[width=\linewidth]{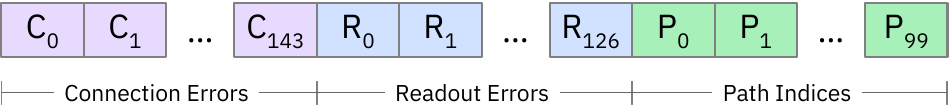}
    \caption{Input Vector, the first 144 features are connection error rates, the next 127 features are readout error rates, and the last 100 features are path indices (visited qubits). Since the size of the input vector needs to be constant but the path length varies, the path is $-1$ padded to the right.}
    \label{fig:input-vector}
\end{figure}

We prepare a feature vector $X$ for a given experiment, including calibration data and normalized path information as described in Fig. \ref{fig:input-vector}.

\subsection{Training}
Each experiment's path and calibration data are encoded into feature vectors as described in Sec. \ref{sec:train-data-preparation}, which are inputs to our gradient boosting model. The gate fidelity measurements associated with each experiment are used as ground truths for model training. The training process is conducted iteratively, adjusting hyperparameters to minimize prediction error and improve model generalization.

\subsection{Validation}
\begin{figure}[h!]
        \centering
        \scalebox{0.55}{\input{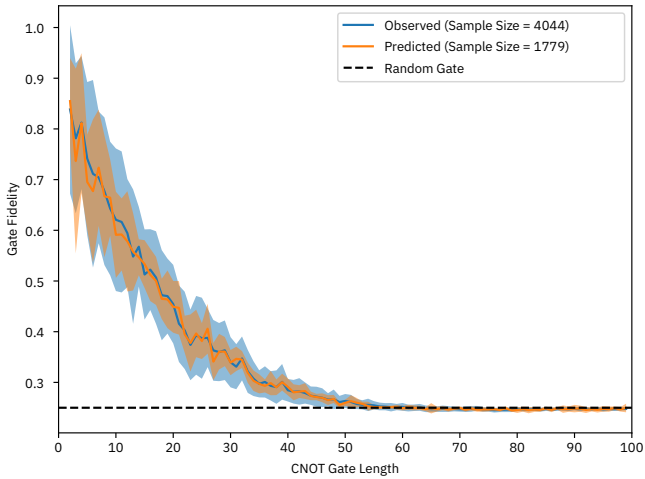}}
        \caption{Gate Fidelity as a function of CNOT Gate Length for i) Experimentally measured on \texttt{ibm\_quebec}, ii) Predicted by XGSwap.}
        \label{fig:gate-fidelity-vs-path-length}
    \end{figure}

For validation, we use a subset of the data--not previously presented to the model for training--to assess the model's performance. 
The validation results captured in Table \ref{tab:training-metrics} underscore the XGBoost model's effectiveness in predicting gate fidelity with a Mean Absolute Error (MAE) of $3.18\times 10^{-2}$, indicating a high level of precision in its predictions. The Mean Squared Error (MSE) and Root Mean Squared Error (RMSE) further reinforce this, with values of $4.8\times 10^{-3}$ and $7.0\times 10^{-2}$, respectively, suggesting that the model's predictions are generally close to the actual values with minimal deviation. The $R^2$ value of 0.87 indicates that approximately 87\% of the variance in the fidelity data is predictable from the features used by the model, signifying a strong predictive power. This is further corroborated by the plot in Fig. \ref{fig:gate-fidelity-vs-path-length}, which shows a close alignment between the model's predictions and the actual gate fidelity measurements.
\begin{table}[h!]
    
    \centering
    \caption{Performance metrics of the XGBoost model on the validation dataset.}
    \scalebox{0.8}{    \begin{tabular}{@{}cccc@{}}
        \toprule MAE & MSE &RMSE &$R^2$ \\  \midrule
        0.0318              & 0.0048             & 0.07                    & 0.87      \\ \bottomrule
        \end{tabular}}

    \label{tab:training-metrics}
    \end{table}

    To further enhance the model's performance, hyperparameter tuning could be employed to optimize settings such as the learning rate and max depth. Additionally, exploring more sophisticated feature engineering or employing advanced ensemble techniques like stacking might improve prediction accuracy and robustness. 
\section{Results}
In this section, we present the results from running a random CNOT gate benchmark on the IBM Quantum device \texttt{ibm\_quebec}. We compare the fidelity of CNOT gates executed using the shortest path with those executed using the path identified by XGSwap. The results provide insights into the efficacy of XGSwap in improving gate fidelity by selecting paths that are not necessarily the shortest but are expected to yield higher fidelity.
\subsection{Experimental Setup}

In this series of experiments, we select pairs of control and target qubits at random and execute CNOT gates along paths between them. The control setup uses the shortest undirected path for gate routing, whereas for the experimental trials, we employ XGSwap to identify the path expected to yield the highest gate fidelity. To estimate the fidelity of the process, we run full process tomography, and the resulting process fidelity is converted to gate fidelity following Eq. \ref{eq:gate-fidelity}. This procedure is repeated more than 300 times for random CNOT gates between 3 and 27 qubits.

\subsection{Results}

We observe that the path identified by XGSwap is different from the shortest path in 138 out of 313 experiments. Tab. \ref{tab:random-cnot-results2} provides a breakdown of the results for these 138 experiments. In 76 out of 138 experiments, the path identified by XGSwap yields higher fidelity than the shortest path.
 \begin{table}[h!]

    \centering
    \caption{Experiments where the path identified by XGSwap is different from the shortest path. Experiments on \texttt{ibm\_quebec} in Mar' 2024.}
    \label{tab:random-cnot-results2}
    \scalebox{0.8}{\begin{tabular}{@{}lcc@{}}
        \toprule
        Metric                               & Number of Experiment & Proportion\\  \midrule
        Experiments                  & 138           & 100\%       \\ \midrule
        Lower Fidelity   & 62        & 45\%           \\
        Higher Fidelity  & \textbf{76}     & \textbf{55\% }            \\ \bottomrule
        \end{tabular}}
    \end{table}
 
\begin{figure}[h!]
    \centering
    \includegraphics[width=\columnwidth]{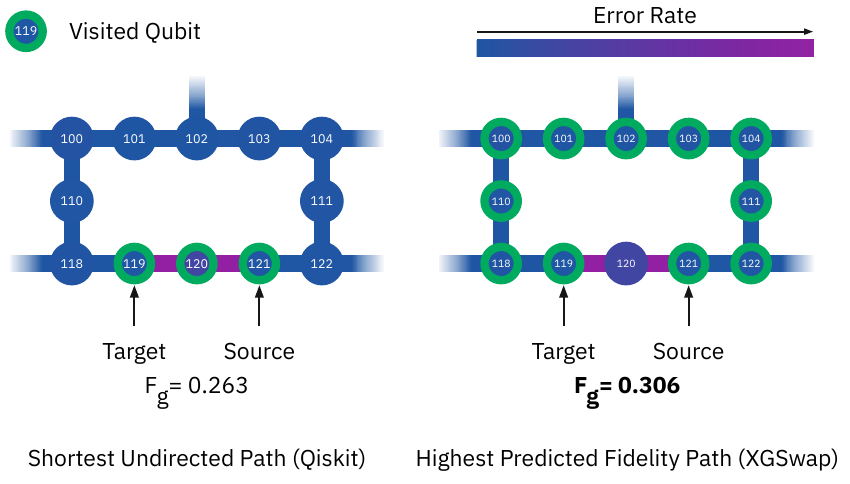}
    \caption{Coupling Map drawing for a case where the XGSwap leads to a higher fidelity. The shortest path on the left uses two high error rate connections between qubits 119, 120, and 121. XGSwap correctly identifies and routes around these connections.}
    \label{fig:better}
\end{figure}

One example where XGSwap predicts a higher fidelity path is shown in Fig. \ref{fig:better}, where the shortest path uses two high error rate connections between qubits 119, 120, and 121. For those two connections, the calibration returns an error rate of $1$ which likely means a failed calibration. XGSwap correctly identifies and routes around these connections, using an average connection error of $1.4 \pm 0.7 \times 10^{-2}$, resulting in higher fidelity.

\begin{figure}[h!]
    \centering
    \includegraphics[width=\columnwidth]{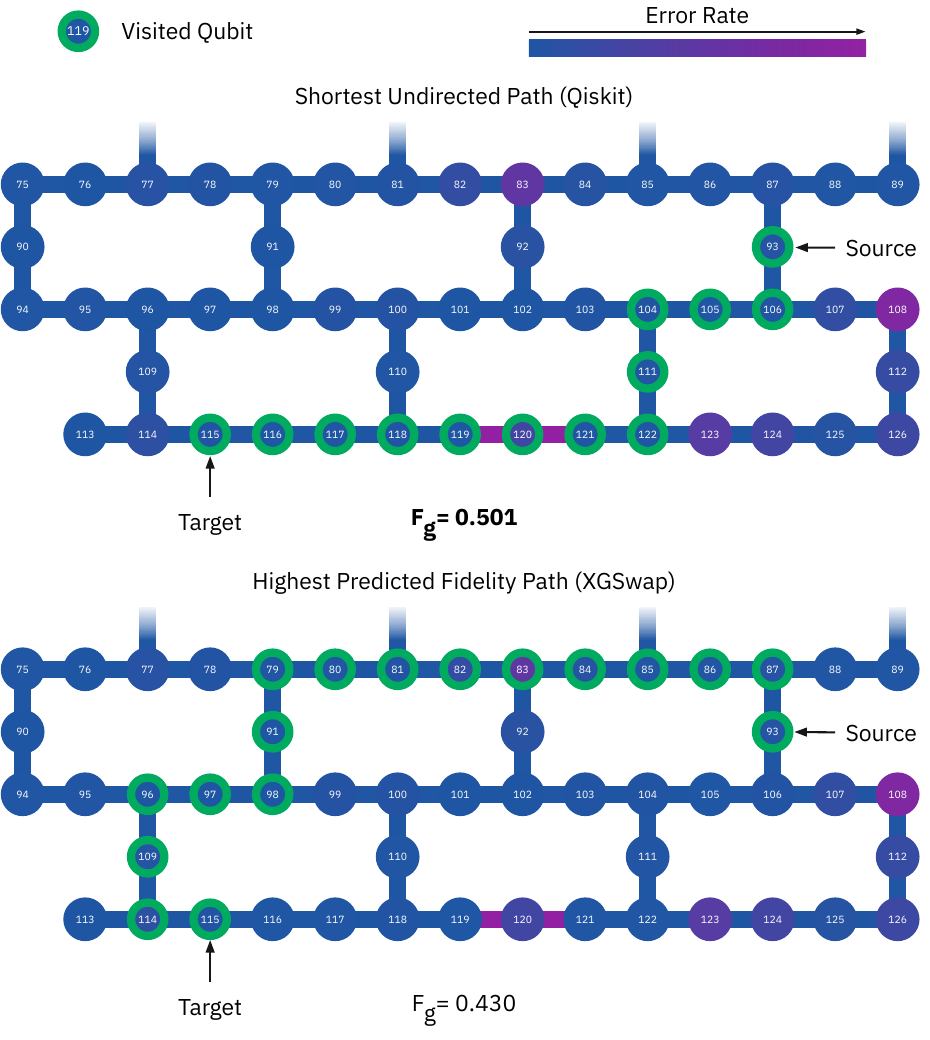}
    \caption{Coupling Map drawing for a case where the XGSwap leads to a lower fidelity. The shortest path on the left uses two high error rate connections between qubits 119, 120, and 121. XGSwap correctly identifies and routes around these connections but yields lower fidelity.}
    \label{fig:notasgood}
\end{figure}

However, there were also instances where the paths chosen by XGSwap did not result in higher fidelity. For example, in Fig. \ref{fig:notasgood}, the shortest path uses two high error rate connections between qubits 119, 120, and 121. In this case, the connection error is, on average, $1.7 \pm0.37 \times 10^{-1}$. XGSwap correctly identifies and routes around these connections, using connections with an average error of $1.3 \pm0.33 \times 10^{-1}$, yet the fidelity is lower. Additionally, there is evidence suggesting that this result, may be partially attributed to outdated or failed calibrations. Indeed, a connection error of $1$ such as the one observed between qubits 119, 120, and 121, should lead to a fidelity value of $1/4$, that is the fidelity of a random gate. However, this was not observed in our results and suggests that some calibration data is incorrect.

\subsection{Discussion}
The results from our random CNOT gates benchmark provide early insights into the practical efficacy of employing a gradient boosting based routing strategy over the traditional shortest path method. Notably, in approximately 44\% of cases, XGSwap identified a path that deviated from the shortest path. This different path led to an increase in fidelity in about 24\% of the total experiments, suggesting that in certain scenarios, the shortest path is not the most fidelity-efficient. These results support the hypothesis that quantum routing can significantly benefit from machine learning models that predict and optimize for higher fidelity, taking into account the complex noise characteristics of the target quantum system.
Furthermore, cases where the fidelity was higher or lower even though the path was the same, show that factors other than the routing path (such as slight variations in qubit performance over time) can influence fidelity outcomes and should thus be discarded.

\section{Conclusion}
In this paper, we introduced XGSwap, a gradient boosting model designed to predict the fidelity of CNOT gates in various routing paths, thereby addressing the complex challenges presented by the coupling constraints of quantum processors. 
This model, trained on an extensive dataset of over 4000 random CNOT paths of varied lengths, demonstrated its potential by identifying higher fidelity paths in about 24\% of the cases, showing that longer paths can sometimes lead to better quantum gate performance than the shortest possible routes.
However, our model solely predicts the best path for a given CNOT gate, and does not consider the overall circuit. This means that the model may not always provide the best routing solution for the entire circuit. Better outcomes could be reached by using more sophisticated machine learning models, such as deep learning, and not limiting the model to predicting the best path.
Lastly, the model was trained on a single quantum processor, and we can only speculate on its performance on other devices.
\section*{Acknowledgements}

We would like to thank PINQ2 for the access to the machine \texttt{ibm\_quebec} and the computation time needed for this study. 

\bibliographystyle{IEEEtran}
\bibliography{references}

\end{document}